\documentstyle{laa}
\begin{document}
\thesaurus {11.05.2, 11.07.1, 11.09.2, 11.16.1, 11.16.2, 11.19.2}

\title{ Tidally-triggered disk thickening}
\subtitle { II. Results and interpretations}
\author{Vladimir Reshetnikov\inst{1,2}, Fran\c coise Combes\inst{2}}

\offprints{F.Combes, bottaro@obspm.fr}

\institute{ Astronomical Institute of St.Petersburg State University,
           198904 St.Petersburg, Russia
\and
            DEMIRM, Observatoire de Paris, 61 Av. de l'Observatoire,
 F-75014 Paris, France}

\date{Received 1996; accepted}

\maketitle

\begin{abstract}
 We have reported in a previous paper (Paper I) $B$,$V$ and $I$ band photometric 
data for a sample of 24 edge-on interacting spiral galaxies, together
with a control sample of 7 edge-on isolated galaxies. We discuss
here the main result found in this study: the ratio $h/z_0$ of the
radial exponential scalelength $h$ to the constant scaleheight $z_0$
is about twice smaller for interacting galaxies. This is found to be
due both to a thickening of the plane, and to a radial stripping
or shrinking of the stellar disk. 
If we believe that any galaxy experienced a tidal interaction in 
the past, we must conclude that continuous gas accretion and
subsequent star formation can bring back the ratio $h/z_0$ to
higher values, in a time scale of 1 Gyr.

\keywords{ galaxies: evolution, general, interactions, photometry, peculiar, 
spiral }

\end{abstract}

\section{Introduction}

>From the seminal work of Toomre \& Toomre (1972) we know how sensitive
are galaxy disks to tidal interactions, from the formation of tidal
tails and bridges up to the complete disruption of initial disks
into a merged system, looking as an elliptical. Even non-merging 
interactions or minor mergers can thicken and destroy a stellar disk, and this
has been advanced as an argument against frequent interactions in
a galaxy life, or formation of the bulge through minor mergers in spiral 
galaxies (e.g. Gunn 1987). Although many vestiges of interactions
are observed in present-day galaxies (ripples, shells, plumes, 
Schweizer 1990), it was assumed that these interactions must have
occured early in the galaxy life, before the formation of the disk.

  Toth \& Ostriker (1992) have used the argument of the fragility of
disks to constrain the frequency of merging and the amount of
accretion, and draw the implications on cosmological parameters.
They claim that the thickness of the Milky Way disk
implies that no more than 4\% of its mass can have accreted
within the last $\rm 5\,10^{9}$ yrs; moreover they question the currently
fashionable theory of structure growth by hierarchical merging, which 
would not be supported by the presence of thin galactic disks, cold
enough for spiral waves to develop. 

Numerical simulations have been performed to check how fragile 
galactic disks are with respect to interaction and merging
(Quinn et al 1993, Walker et al 1995). But the results depend strongly
on the gas hydrodynamics and star-formation processes, since the 
thin disk can be re-formed continously through gas infall.
The gas accreted can, by its dissipative
character, settle in a thin plane, and make the stellar disk thinner
by its direct gravitational effect, and by the subsequent
star formation.
 Other unknown parameters could influence the results, as the
actual efficiency of dynamical friction, the true flattening of 
the dark matter, etc...
Since the question is not completely settled, it is interesting 
to check observationally the influence of tidal interactions on
the thickening of galaxy disks.

In a preliminary study, Reshetnikov et al (1993)
have shown that the disks of strongly interacting spirals are 2-3
times thicker as compared with the disks of normal spiral galaxies.
But the sample was only made of 6 galaxies and was too small to derive
statistically significant results.
Here we investigate the efficiency of tidal disk thickening by
observing the thickness of planes in a large enough sample
of edge-on interacting galaxies, in comparison to a control
sample of isolated galaxies. In Paper I, we have presented
the observations and general photometric results.
In this report, after describing the reduction methods 
(section 2), we analyse the thickness of the planes,
at various radii, show that it remains almost constant
in a given galaxy, i.e. stellar disks are not
strongly affected by warping and flaring (section 3).
 Then we study the statistical variations of the ratio $h/z_0$ of
the radial scalelength $h$ to the scaleheight $z_0$,
and discuss the results in section 4.

\section{Observations and data reduction}

\subsection{Sample and observations}

For a statistical study of the influence of galaxy interactions
on the $z$-structure of the disks of involved galaxies, we
selected a sample of apparently edge-on spiral galaxies
belonging to strongly interacting systems. Our sample consists of 24
interacting systems containing at least one edge-on galaxy.
The sample is relatively complete since we included
practically all known interacting systems that are suitable
in angular diameter and could be observed during our observational
run (see Paper I). 
All interacting systems (as far as 7 isolated galaxies) were
observed at the OHP 1.2 m telescope in the $B$, $V$ and $I$
passbands. General photometric results of the observations
(including isophotal maps of all objects) are presented in
Paper I.

\subsection{Reduction}

We considered photometric cuts (4-6 typically) along 
minor axes of the sample galaxies at different galactocentric 
distances. For the galaxies with warped disks the directions of
the minor axes were determined locally as a perpendicular to
local major axis direction.
The positions of the galaxy planes were determined
by averaging the vertical profiles and under assumption of
symmetrical light distributions with respect to the planes.
In general the vertical cuts of interacting galaxies look 
quite symmetric (especially in the $I$ passband). 
Comparing asymmetry of the averaged profiles of interacting
galaxies with asymmetry of the minor axis profiles in our
sample of isolated spirals, we concluded that interacting and 
normal galaxies are distributed approximately in the same 
range of inclinations with respect to the line of sight. 
Following Guthrie (1992), we found that our sample isolated
galaxies are, on average, within $\rm 6^{o}$ from edge-on orientation 
and, therefore, concluded that most interacting edge-on galaxies 
are in the same range also. Let us note also that according to
van der Kruit \& Searle (1981a) and Barteldrees \& Dettmar (1994)
moderate (5$^{o}$-10$^{o}$) deviations from edge-on orientation do not 
significantly change
the slope of vertical surface brightness distribution.  
Moreover, the control  sample of edge-on isolated galaxies is also 
contaminated by not exactly edge-on objects, so this systematic effect
is somewhat compensated when both samples -- interacting and
non-interacting -- are compared.

By statistically studying the disk thickness of edge-on 
interacting galaxies, we did not consider question about best
fitting function for each galaxy and fitted all the averaged 
vertical profiles 
by standard $sech^2(z/z_{0})$ law (van der Kruit \& Searle
1981a), where $z$ is the distance from the galaxy plane and
$z_{0}$ is the scale height. (At large $z$ a comparison 
between $z_{0}$ value and exponential scale height $h_{z}$
is possible via $z_{0}~=~$2$h_{z}$.) We choose this function
in order to have the largest possible comparison sample of normal
galaxies with uniformly determined scale heights. We found that 
galaxies with published $z_{0}$ values are predominant
in the literature (e.g., van der Kruit \& Searle 1981a,b and 1982a,b
(vKS); Barteldrees \& Dettmar 1994 (BD)). It should be noted
also that ``$sech^{2}$-distribution'' gives quite satisfactory
(within $\rm 0.^{m}2$)
approximation for most of the sample. 
This can be understood, since
this is the distribution of a self-gravitating isothermal layer
of stars; and here the stellar component is representing most
of the mass (the gas mass is negligible, and the dark halo
mass is very small within the optical disk), and there is
only small departures from z-isothermality (e.g. van der Kruit
1988). Close to the plane, the stars are cooler, but the vertical
dispersion at worst can be represented by a $sech(z)$ vertical
density profile (instead of a $sech^2(z)$, cf Bottema 1993).

We excluded central regions of the galaxies from
our study in order to avoid the bulge light contribution to
the fitted profiles. After inspection of the radial
surface brightness distribution, we considered vertical cuts
at radii along the major axis where the bulge influence is
negligible. From the other side, in order to have reasonable 
photometric profiles with amplitude (difference between
central surface brightness and faintest level of the cut)
larger than $\rm 2^{m}$
we excluded faint outer regions of the galaxies.
We find that our vertical cuts are distributed, on
average, between 1 and 2.4 exponential scalelengths $h$ (see below)
of the galaxy disks in the $I$ passband.

For some of our sample galaxies the seeing is rather bad (larger than
2\arcsec). The seeing effects do not strongly affect the slopes of
the surface brightness profiles (see, for instance, Nieto et al. 1990)
and cannot noticeably influence our results. To check the size
of this effect, we compared scale height values determined
from the original images of the galaxies and after
Lucy-Richardson restoration (for this we
used standard MIDAS routine, the PSFs were constructed from the
star images in the corresponding frames). We found that
original frames give about 20\% systematically larger values
of scale heights for the thinnest (in comparison with star images)
galaxies. For most of our objects the effect of seeing is
insignificant. Therefore, we corrected $z_{0}$ values of some flat
galaxies (with $z_{0}$ about or less 2\arcsec)
for this effect (correction of at most 20\%).

To study the structure of galaxies in the radial direction, we
extracted major axis profiles of all galaxies. Excluding central
bulge-dominated regions, we determined exponential scalelengths
by fitting outer parts of the profiles. The $B$ band radial cuts
of our sample interacting galaxies often look peculiar and
asymmetric and does not provide reasonable fit. Therefore,
we will use the $I$ band derived scalelengths in the following discussion
(photometric profiles in the $I$ filter are significantly smoother and
more regular). Let us noted also that our exponential fit
of the sample galaxies does not show any systematical
difference with literature data. For instance, average central surface
brightness of the disks in the $B$ band ($\mu_{0}(B)~=\rm~21.2\pm0.6$)
is close to the canonical Freeman's (1970) value. The mean $B$ to $I$
scale length ratio ($\rm 1.18\pm0.26$) is in good agreement
with de Grijs \& van der Kruit (1996) (dGvK) value ($\rm 1.17\pm0.10$). 

\section{Results}

General characteristics of our sample edge-on galaxies are presented
in Table 1. The columns of the table are: galaxy name (see Paper I for
galaxy identification); morphological type according to 
NED\footnote{The NASA/IPAC Extragalactic Database (NED) is operated
by the Jet Propulsion Laboratory, California Institute of
Technology, under contract with the National Aeronautics and Space
Administration.}; adopted distance ($H_{0}=$75 km/s/Mpc); 
corrected for Galactic absorption absolute
blue luminosity and colour $B-V$
of the galaxy (apparent magnitudes and colours of galaxies are
from Paper I); 
total HI mass, in $\rm 10^{9}~m_{\sun}$, according to
Huchtmeier \& Richter (1989), de Vaucouleurs et al (1991) (RC3)
and LEDA (for the members of Arp 242 we used associated with the
central galaxies HI masses according to Hibbard \& van Gorkom 1996);
HI mass-to-blue luminosity ratio, in $\rm m_{\sun}$/$L_{\sun}^{B}$
(adopting $M_{\sun}^{B}=+$5.48); 
average scale height in the $I$ 
passband; scalelength to
scale height ratio in the $I$ passband. For two objects - Arp 208 and
UGC 11230 - we found no available redshifts and present in Table 1
scale height values in arcseconds.

Two galaxies - Arp 124NE and Arp 127S - are not presented in the table.
Arp 124NE demonstrates notable deviation from edge-on orientation
(we see dust lane shifted from the galaxy nucleus, photometric
profiles along the minor axis look very asymmetric). Arp 127S is
embedded in a relatively bright extended envelope which contribute
significantly to the observed surface brightness distribution along
the minor axis. Moreover, according to NED the components of 
Arp 127 demonstrate very different radial velocities (Arp 127N - 
5065 km/s, Arp 127S - 13650 km/s) so the nature of this 
double system is unclear.

The distribution of the scale height values in three passbands
as a function of position
along the major axis is shown in Fig.1 for all sample galaxies
with more than two measurements of $z_{0}$.

\begin{figure*}
\picplace{1 cm}
\caption[1]{The scale height distributions for the sample galaxies as
a function of radius along the major axis. Both axes are in kpc (in
arsec for Arp 208 and UGC 11230). Circles represent
the data in the $I$ passband, squares - $V$, stars - $B$. The figures
orientations is such that the distance increases from S to N or from 
E to W. We use general name of interacting system for the systems with
one edge-on galaxy and give more detailed name for the objects with
less clear identification (see Fig.1 in PaperI).}
\end{figure*}

\begin{table*}
\caption[1]{General characteristics of the sample galaxies}
\begin{tabular}{lllllllll}
\\
\hline \\
Galaxy & Morphological & Distance & $-M_{B}$ & $(B~-~V)_{\rm 0}$ & m(HI) & 
m(HI)/$L_{B}$ & $z_{0}(I)$ & $h/z_{0}(I)$  \\
       & type          & (Mpc)    &    &  & ($\rm 10^{9}~m_{\sun}$) &  &
(kpc) &  \\ 
\\
\hline \\
Arp 30N & dm & 113.6 & 19.89 & 0.43 & 3.7 & 0.26 &1.62 & 2.2 \\
Arp 71A & c & 148   & 20.89  & 0.90 & 7.7 & 0.22 &2.40 & 5.0 \\
Arp 112E&   &  70.4 & 17.94 &0.60 &1.9 & 0.81   &1.31 & 1.4 \\
Arp 121SW& a & 76.3 & 20.40 &1.01 & & &1.26 & 3.4 \\
Arp 150SE&   & 159  & 21.09 &0.89 & & & 1.65 & 2.1 \\
\hspace*{1.27cm}SW& m & & 20.40 &0.77 & & &1.35 & 3.2 \\
Arp 208W &   &      &    &0.62 & &  & 1.2\arcsec & 4.8 \\               
Arp 242N &   & 88.7& 20.01 &0.81 & 1.1 & 0.07 & 2.33 & 1.8 \\
\hspace*{1.27cm}S & O/a & & 20.28 &0.82 &1.3 & 0.06 &2.09: & 1.0: \\
Arp 278NW&  & 67.2  & 20.26 &0.63 &11.3 & 0.57 &1.38 & 4.0 \\
\hspace*{1.27cm}SE& & & 19.80 &0.55 &6.1 &0.47   &1.56 & 2.5 \\
Arp 284E & m & 40.1  & 18.49 &0.46 &4.7 &1.21 &1.46 & 3.2 \\
Arp 286 & b & 22.7  & 17.94 &0.85 &0.76 &0.33&0.76 & 3.3 \\
Arp 295SW & c & 94.0  & 20.24 &0.91 &10.0 & 0.52&1.42 & 2.7 \\
VV 426SE  & c & 67.2  & 18.51 &0.34 &7.6 & 1.93 &1.55 & 1.3 \\
VV 489S   &   & 100.4 & 20.33 &0.52 & & &1.22 & 3.7 \\
VV 490N   & c & 97.9  & 20.35 &1.11& & &2.48 & 2.3 \\
\hspace*{1.2cm}S& &  & 19.28 &1.03& &  &1.03 & 1.3 \\ 
VV 679A   &   & 59.9 & 19.89 &0.54 &10.7 & 0.76 &1.75 & 3.2 \\
VV 773E   &   & 74.2 & 20.77 &-0.28& & &1.31 & 2.2 \\
\hspace*{1.2cm}W& &  & 19.04 &-0.09& &  &0.73 & 3.1 \\
K 3W      & c & 99.3 & 18.70 &1.00& 11.1& 2.36 & 1.14 & 6.4 \\
K 10E     & c & 30.3 & 18.77 &0.51& 7.7 & 1.54 &1.00 & 2.3 \\
\hspace*{0.75cm}W & & & 16.80 &0.48& & &0.44 & 1.6 \\
K 14SW    &   & 61.6 & 19.13 &0.88& & &1.25 & 4.5 \\
K 540NE   &   & 82.2 & 19.42 &0.33& & &0.87 & 3.7 \\
K 547E    & c & 105.5& 19.82 &0.85& & &1.71 & 2.2 \\
K 585     & b & 79.9 & 19.29 &0.66& 18.2& 2.25 &1.54 & 3.3\\ \\
UGC 11132 & b  & 41.0 & 18.37 &0.78& 3.5& 1.02 &0.60 & 4.2 \\
UGC 11230 & c  &      &       &0.79& &   & 3.2\arcsec & 6.1 \\
UGC 11301 & c  & 62.8 & 20.58 &0.50& & &0.47 & 19: \\
UGC 11838 & d  & 50.1 & 18.65 &0.52& 5.75 & 1.28 &0.85 & 6.5 \\
UGC 11841 & cd  & 83.8 & 20.09 &0.93& & &1.86 & 7.2 \\
UGC 11859 & c  & 42.8 & 17.86 &0.75& 6.1 & 2.82 &0.45 & 6.6 \\
UGC 11994 & c  & 68.7 & 20.13 &0.90& 12.4 & 0.71 &1.16 & 3.5 \\
\\
\hline
\end{tabular}
\end{table*}

\subsection{Scale heights}

\subsubsection{Radial distribution of $z_{0}$}

It is well established that the scale parameter $z_{0}$
is almost independent of radial distance for the disks of normal
spiral galaxies (vKS, Shaw \& Gilmore 1990, dGvK). Our results
for seven non-interacting late-type spiral galaxies confirm
this conclusion. As one can see in Fig.1, normal spirals
demonstrate small variations of scale height with radius.
We find that the mean dispersion of $z_{0}$ values in
the $I$ band is 7\%$~\pm~3$\% only (note that this value
refer to relatively bright region of the galaxy disk - see
item 2.2). This constancy level is consistent with the level
of 10-15\% found in dGvK. 

In agreement with literature data (e.g., Shaw \& Gilmore 1990,
dGvK), we did not find $z_{0}$ variations
with passband for our sample isolated spiral galaxies.
The mean ratio of scale heights in the $I$ and $B$ passbands
is $\rm 1.04~\pm~0.07$ for seven isolated spiral galaxies.
 
Visual inspection of Fig.1 shows that interacting spirals
demonstrate more variety in behaviours than isolated galaxies.
The ratio of scale height dispersion to the mean value
of $z_{0}$ is equal to $\rm 11~\pm~6$\% for them in the $I$
passband. This is almost twice as for non-interacting galaxies.
The distribution of ${\rm \sigma}(z_{0})/<z_{0}>$ values for
our sample interacting and isolated galaxies is compared in 
Fig.2. The distribution for interacting spirals
is significantly wider and about half of interacting spirals
demonstrate larger scale height variations than it was found
for isolated spirals.

\begin{figure}
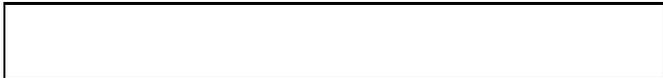

\picplace{1 cm}
\caption[2]{Distributions of relative variations of scale height
for isolated (dashed line) and interacting (solid line) galaxies in 
the $I$ passband.}
\end{figure}

The most frequent feature of radial $z_{0}$ behaviour
of interacting spirals is systematical change of scale height
along the disk (see Fig.1). One can attribute about 40-50\% of
the galaxies to such ``slanting'' disks.
Probably. ``slanting'' disk is a transient phase
of galaxy evolution caused by large-scale asymmetry of
the potential due to proximity of massive companion.
Four galaxies (Arp 121, Arp 295, VV 679, and K 14) show 
increase of scale height to the central region. Apparently,
such disk structure reflects the presence of a bar in the galaxies.

Only one galaxy - VV 490N - demonstrate clearly the radial
increase of the disk scale height. Such disk behaviour
is expected for the galaxies subject to accretion of 
small companions (see Fig.1 in Toth \& Ostriker 1992, Fig.6
in Quinn et al 1993, Fig.9 in Walker et al 1996). 
The double system VV 490 resembles remarkably numerical models
studied in those works - it consists of large spiral galaxy
(VV 490N) and small companion (VV 490S) probably settling 
to the plane of the primary (Fig.1 in Paper I). The strong
flaring of the VV 490N disk begins at $\rm r\approx~$2$h$ and
reaches about 50\% at $\rm r\approx~$3$h$. This is in general
quantitative agreement with results of numerical simulations.
One can note also that analogous flaring structure of stellar
disks was found recently in two interacting galaxies - NGC 3808B
and NGC 6286 - subject to strong matter accretion from the 
companions (Reshetnikov et al 1996). 

\subsubsection{Mean scale heights of interacting and isolated galaxies}

The average scale heights in the $I$ passband for the sample
galaxies are presented in Table 1. As for isolated spirals, we
found no colour dependence of $z_{0}$ values for interacting
galaxies also. The mean ratio of scale heights in the $I$ and
$B$ passbands is $\rm 1.00~\pm~0.11$ for 27 galaxies.

We compare mean values of $z_{0}$ for interacting
and normal (``field'') galaxies in the Table 2. 
(Note that according to
Karachentsev et al 1993 two non-interacting galaxies in 
our sample - UGC 11301 and UGC 11841 - are ``Malin 1'' type 
galaxies. We excluded them from further consideration.) 
As one can see, interacting galaxies show larger mean value
of $z_{0}$ in comparison with our and vKS samples of
isolated spirals. The BD sample gives significantly larger
average scale height with large dispersion.

\begin{table*}
\caption[2]{Comparison of the samples}
\begin{tabular}{lllllll}
\\
\hline \\
Sample & $N$ & $-M_{B}$ & Passband & $h$ (kpc) & $z_{0}$ (kpc)  & $h/z_{0}$ \\
\\
\hline \\
IGs (our)    & 29 & $\rm 19.6~\pm~1.0(\sigma)$ & $I$ & $\rm 4.0~\pm~2.4(\sigma)$ &
$\rm 1.43~\pm~0.49(\sigma)$ & $\rm 2.9~\pm~1.2(\sigma)$ \\ 
Field (our) & 5 &   $\rm 19.3~\pm~1.0$ &$I$ & $\rm 3.7~\pm~1.1$& $\rm 0.77~\pm~0.27$ & 
$\rm 5.4~\pm~1.3$ \\
Field (vKS)     & 8 &$\rm 18.6~\pm~1.0$ &$J$ &$\rm 4.5~\pm~1.9$ &$\rm 1.06~\pm~0.51$& 
$\rm 4.6~\pm~1.6$ \\
Field (BD)      & 19&$\rm 19.2~\pm~0.7$ &$g,r$&$\rm 6.6~\pm~2.2$ &$\rm 1.67~\pm~0.76$& 
$\rm 4.4~\pm~1.5$ \\
Field (dGvK)    & 8 &$\rm 18.2~\pm~2.0$ &$I$ & $\rm 3.25~\pm~1.3$& &
$\rm 5.9~\pm~0.4$ \\
\\
\hline
\end{tabular}
\end{table*}

The cumulative distribution of scale height values for normal galaxies
(we summarized our, vKS and BD samples) is compared with 
distribution for interacting spirals in Fig.3a,b. Both
distributions show the same widths with global peaks at
1.5 kpc for interacting galaxies and about 1 kpc for field
spirals. 
Trying to understand the origin of the secondary
peak in Fig.3a at $z_{0}\geq$2 kpc, we inspected normal
galaxies falling in this region. Using the DSS\footnote{The
Digitized Sky Surveys were produced at the Space Telescope
Science Institute under U.S. Government grant NAG W-2166.} 
images, we found that among
8 normal galaxies with $z_{0}\geq$2 kpc (7 of them are from
the BD sample) at least 6 have comparable size companions
within 5 optical diameters. Moreover, several galaxies 
demonstrate signs of significant non-edge-on orientation. 
For instance, the most striking galaxy in the BD sample -
ESO 460-G31 - with $z_{0}~=~$3.9 kpc has a companion at
one optical radius and noteably shifted from the nucleus 
dust lane. Therefore, the secondary peak at Fig.3a
consists of non-isolated galaxies mainly. Normal non-interacting
galaxies possess disks with scale height about 1 kpc.
This conclusion is in agreement with results obtained from
the dGvK sample. According to de Grijs \& van der Kruit (1996),
the exponential scale height of thick disk of spiral galaxies
range from about 470 pc to 620 pc. Transforming exponential
scale heights to $z_{0}$, we have range 0.94-1.24 kpc that
is in good agreement with Fig.3a distribution.
 
\begin{figure}
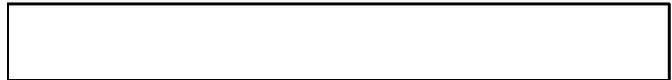

\picplace{1cm}
\caption[3]{Distribution of scale heights (in kpc) for
normal (a) and interacting (b) galaxies.}
\end{figure}

We did not take into consideration possible biases due to
different morphological composition and luminosity
distribution of the samples of normal and interacting galaxies
in our previous analysis.
As one can see in Table 1, only half of interacting galaxies
have estimated morphological types. Within limits of poor
statistics, the distributions of morphological types
of our sample edge-on interacting galaxies and normal galaxies 
in cumulative sample (our+vKS+BD) are undistinguishable.
Moreover, both samples do not show statistically significant
correlations of scale height on morphological type.
>From the other side, the absolute luminosity distributions
in both samples are close also (for instance, mean absolute
blue luminosities of the samples galaxies are -19.1 and
-19.6 for normal and interacting galaxies correspondingly with
dispersion about $\rm 1^{m}$). 

In Fig.4 we compare distributions
of normal and interacting galaxies in the plane absolute blue
luminosity $-$ scale height. As one can see, both samples 
occupy approximately the same region in this plane.
Solid line in Fig.4 shows $ L \propto z_{0}^{2}$ dependence
expected for normal face-on galaxy with $\mu_{0}(B)~=~$21.65
(Freeman 1970) and $h/z_{0}~=~$5 (vKS, Bottema 1993).
The data for real edge-on galaxies follow this relation
quite acceptably with some systematic shift. This shift is
a measure of total absorption in the disk of edge-on galaxy
in comparison with face-on orientation. Considering only
bulgeless galaxies, we find an estimation of total
absorption in the $B$ band as $\rm 1.6~\pm~0.2$ mag. This
estimation is in remarkable agreement with the RC3 
value of 1.5 mag.   

\begin{figure}
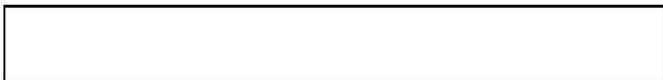

\picplace{1cm}
\caption[4]{Distribution of normal (circles) and interacting
(solid triangles) edge-on spiral galaxies in the plane absolute
blue luminosity $-$ scale height. Solid line represents expected
dependence for the ``standard'' (see text) face-on galaxy.}
\end{figure}

Summarizing the results of direct comparison of $z_{0}$
distributions for normal and interacting galaxies, one
can conclude that there is an evidence of moderately
enhanced (at about 50\%) disk thickness in interacting 
spirals.
 
\subsection{$h/z_{0}$ ratio}

One can expect that normalized thicknesses - $h/z_{0}$ ratios -
can give more distinct evidence of enhanced thicknesses
of interacting disks than absolute values of scale heights. 
Indeed, as one can see in Table 2, interacting galaxies
demonstrate 1.5-2 times lower mean value of $h/z_{0}$
in comparison with all considered samples of normal spirals.

The distribution of $h/z_{0}$ ratios for normal galaxies in
the joint sample (our+vKS+BD+dGvK) is compared with distribution
for interacting spirals in Fig.5a,b. (We neglect in this figure
the possible colour dependence of the $h/z_{0}$ ratios. But note that
exclusion of the vKS data obtained in a relatively blue (close
to the $B$) passband does not change the general shape of distribution
for normal galaxies.) 
As one can see in the figure, both samples demonstrate remarkably
different distributions - normal galaxies are peaked at
$h/z_{0}~\approx~$4-5 while interacting spirals show concentration 
around 2-3. Analysing with the DSS spatial environment of non-interacting 
galaxies having $h/z_{0}~\leq~$3 (left wing of distribution in Fig.5a), we 
found that among 7 such galaxies 5 are non-isolated (see also item
3.1.2). From the other side, among 5 interacting spirals with
$h/z_{0}~\geq~$4, 3 (Arp 208W, K 3W, and K 14SW) demonstrate
very regular symmetric optical morphology (see Fig.1 in Paper I) 
indicative of relatively weak interaction with companions.
Exclusion of such contaminated galaxies makes the difference
between two distributions in Fig.5 significant at any level of confidence.

\begin{figure}
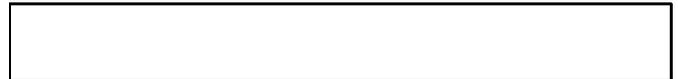

\picplace{1cm}
\caption[5]{Distribution of scalelength to scale height ratios
for normal (a) and interacting (b) galaxies.}
\end{figure}

In Fig.6 we compare scalelength distributions of normal galaxies
in the joint sample and of our interacting spirals. 
Both distributions are statistically undistinguishable although
one can note a somewhat shorter, on average, disks in interacting 
galaxies. Therefore, taking into account
that interacting and normal galaxies demonstrate
statistically undistinguishable distributions of morphological types 
and absolute luminosities also, we can conclude that the difference 
in $h/z_{0}$ ratio must reflect real thickening of interacting disks.

\begin{figure}
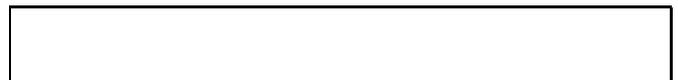

\picplace{1cm}
\caption[6]{Distribution of scalelengths (in kpc) 
for normal (a) and interacting (b) galaxies.}
\end{figure}

\subsection{General observational conclusions} 

In our analysis of the vertical structure of edge-on
interacting galaxies we found that strong tidal influence
increases significantly scale height variations along
galactic disks even within relatively bright parts of them.
The most typical feature of interacting disks is
systematical change of $z_{0}$ along the radius.

>From the direct comparison of scale height and
$h/z_{0}$ distributions in the samples of interacting and
isolated spirals we found an evidence of moderate
(in 1.5-2 times) thickening of galactic disks in
interacting systems. This thickening refer to
the region of exponential disk between 1 and 2.4 of
exponential scalelength (or between 0.6 and 1.4 of
effective radius).

The mean characteristics of edge-on interacting galaxies
in our sample are: absolute blue luminosity $M_{B}~=~\rm 
-19.6~\pm~1.0$ (so ``face-on'' magnitude must be about $\rm
-21$), exponential scalelength $h~=~\rm 4~\pm~2$ kpc.
Therefore, typical galaxy in our sample is comparable with the
Galaxy and M 31. Most edge-on galaxies have comparable
luminosity companions within one optical diameter (see Fig.1
in Paper I). One can conclude from this that tidal interaction
between large spiral galaxies like the Milky Way and the
Andromeda galaxy at the mutual distance about one optical
diameter leads to 1.5-2 times thickening of their stellar
disks at half-mass radius of the luminous disk.

\section{Discussion}

 The principal result of our study is that the ratio $h/z_0$ 
of the radial scalelength $h$ to the constant scaleheight $z_0$
is 1.5 to 2 times higher in interacting galaxies than in isolated
spiral galaxies. Moreover, the average value found  $h/z_0$= 2.9,
is significantly lower than all values found in the literature: 
Bottema (1993) found that the ratio $h/z_0$ varies with 
total galactic luminosity, from 11 for the faintest galaxies to
5.2 for the most luminous, passing through a local minimum of 5.1 
for intermediate luminosities. Both $h$ and $z_0$ increase
steadily with luminosity, $z_0$ by more than a factor 10.
 Here we found for the same galaxy types, a much lower
ratio $h/z_0$, which means that the tidal interaction has
a strong effect in relative disk thickening.

 First we can note that the effect on the ratio $h/z_0$ is
partly due to the absolute increase of $z_0$ (cf figure 3),
but also to the decrease of $h$ (Fig. 6),
so that the effect becomes quite significant on the ratio
(Fig. 5). (The tendency for galactic disks to be
shorter in strongly interacting systems was mentioned
earlier in Reshetnikov et al 1993.)
This means that the tidal interaction not only
thickens the disk, but also strips its radial extent,
or induces a concentration of the stellar disk.
This is not unexpected, since it is well known that 
tidal interaction triggers the tranfer of angular momentum
outwards, and contributes to the shrinking of the
disks and concentration of the mass. This is due to
strong non-axisymmetric disturbances generated in the disk
by the tidal perturbation, e.g. spirals and bars. 
Gravity torques then produce mass inflow inside corotation,
while some mass is expelled in the outer parts, and
take the angular momentum away (Combes 1996). If there is
a massive dark halo, it acts as a receiver of angular momentum
and helps the visible mass to shrink and condense radially
(Barnes 1988).

 All these perturbations, thickening of the plane, and
radial condensation, leading to a decrease of the ratio $h/z_0$,
must however be transient, and disappear after the interaction 
is over, i.e. on a time scale of one Gyr. Indeed, galaxies
experience many interactions in a Hubble time, and those appearing
isolated now, must have passed through an interaction period,
may be leading to a minor merger. Present galaxies must be
the result of merging of some smaller units, according to
theories of bottom-up galaxy formation, where small building
blocks form first, and large-scale structures virialised
subsequently (e.g. Searle \& Zinn 1978, Frenk et al 1988).
 On the observational side, many clues point to
the high number of galaxy interactions: existence of
shells and ripples in a significant fraction of present early-type
galaxies (Schweizer \& Seitzer 1988, 1992), number of presently interacting
systems extrapolated to earlier times (Toomre 1977). A present
day "isolated" galaxy must have experienced at least several interactions
in the past, and has probably accreted several tens of percents of its
mass in the form of discrete subunits (Ostriker \& Tremaine 1975,
Schweizer 1990). This implies that the $h/z_0$ ratio recovers its high
average value of 5-11 after the interaction. Since the stellar 
component alone cannot cool down, this means that the overall reduction of
thickness should be due to gas accretion. Spiral galaxies possess HI
gas reservoirs in their outer parts, that remain available for star formation.
 Some of this gas can be driven inwards by gravity torques due to a
tidal interaction (Braine \& Combes 1993). Through this increase of
potential well in the plane, the stellar component can react with a 
slightly reduced thickness; but the main consequence will be the 
formation of young stars in a very thin layer. The overall thickness will
be reduced, which can explain our statistical results about $h/z_0$.
  The signature of these interacting events are observed at the present
time in terms of different stellar components in the vertical distribution 
of galaxies: the presence of thin and thick disks. In the Milky Way, the
existence of the thick disk has been known for a long time: from {\it in situ} 
star counts in the direction of the south galactic pole 
(Gilmore \& Reid 1983) and 
high proper motions star surveys in the solar neighborhood (Sandage 
\& Fouts 1987). The age, metallicity and kinematics of the stars in the thick
disk are intermediate between that of the thin disk and halo. From the
chemical evolution, it is recognized that a substantial delay must have
occured between the formation of the thick and thin disks (e.g. 
Pardi et al 1995). Recently, Robin et al (1996) determined the main
characteristics of the thick disk population using photometric star counts
and a model of population synthesis. They found a true discontinuity
between thin and thick disks, with no kinematic or abundance gradient in the
thick disk, favoring the merging event hypothesis for the thickening
of the disk. They claim a scaleheight of 760 pc for the thick disk, with 
a scalelength of 2.8 kpc, giving a $h/z_0$ ratio of 3.7, still too small
for a non-interacting galaxy. This might suggest that the interactions
with the many dwarfs companions of our Galaxy (Ibata et al 1994) or 
the Magellanic Clouds have still an action on its thickness. This is even more
obvious with the scaleheight parameters adopted by Reid \& Majewski (1993),
for whom $z_0$ = 1.5 kpc. 
This view is also in agreement with recent chemical models of the
Milky Way. Chiappini et al (1997) show that the recent constraints,
including metallicity distribution of G-dwarf stars, impose that
the disk of the Galaxy has been built at least in two main infall
episodes. The halo and thick disk have formed rapidly (in a time-scale
of 1 Gyr), while the thin disk is much slower to form (time-scale
8 Gyr), and the gas forming the thin disk must come from the
intergalactic medium (and not from the gas shed by the halo). 

As it is evident from the above discussion, one can expect
a dependence between galaxy thickness and global content of
HI. Fig.7 presents distributions of normal galaxies with known
HI mass in the joint sample (our+vKS+BD) and of our sample interacting
galaxies in the planes m(HI)/$L_{B}$ $-$ $z_{0}$ and $h/z_{0}$
($L_{B}$ values corrected for Galactic absorption only.).
There is good ($\rm r=-0.7$) inverse correlation between the
HI content of a normal galaxy and its scale height (Fig.7a).
Relative thickness of a galaxy - $h/z_{0}$ - also correlates
with HI content (Fig.7b). (Also, a related dependence between
$(B-V)_{0}$ colour and galaxy thickness is present but with less
confidence.)
Dashed lines in figure 7 show double regression fits for normal
spirals: $z_0$(kpc) = 0.84 x [M(HI)/L$_B$]$^{-1/2}$ and $h/z_0$ = 5.0 x
[M(HI)/L$_B$]$^{1/2}$, where M(HI) is the total HI mass (in M$_\odot$)
and L$_B$ is the total luminosity of the edge-on galaxy (in L$_B^\odot$)
uncorrected for internal absorption. It
is interesting that the Milky Way is also satisfying the above relations.
Adopting for the absolute luminosity of "edge-on" Milky Way
$M_B \approx$ -20.5 +1.5 = -19.0 and M(HI) = 4 10$^9$ M$_\odot$, we
obtain from the above correlations $z_0$ = 1.0$\pm$0.27 kpc and $h/z_0$
= 4.0$\pm$1.7. These values are in agreement with current estimates
of the Milky Way parameters (e.g. Sackett 1997). 

Interacting galaxies in general follow 
the same relations as normal galaxies but with larger scatter.
Existence of this correlation can be attributed to the 
dissipative character of the gas: after a perturbation event
that heats both the stellar and gas disks, the gas can
dissipate the extra-kinetic energy away, and flatten again
to its regulated thin disk. Regulation is probably due to
gravitational instabilities, as developed by Lin \& Pringle (1987):
instabilities heats the medium until Toomre $Q$ parameter
is high enough to suppress gravitational instabilities. 
 The gas then cools down until instabilities enter
again the process. This feed-back mechanism could explain
why the observed gas velocity dispersion is constant with
radius (e.g. Dickey et al 1990). In very gas rich galaxies,
after a galaxy interaction that has heated the stellar disk,
gas dissipation can quickly makes the galaxy recover its
equilibrium thickness, through star formation in the thin gas
disk. The mass ratio between the thick and thin stellar disk
depends strongly on the gas content, and will determine the
final global disk thickness. Note that this correlation is
also related to that found by Bottema (1993) as a function
of mass. Fainter objects correspond to
 late-type galaxies which are proportionally more gas-rich,
and show the higher $h/z_{0}$ ratios. 

\begin{figure}
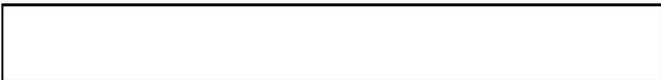

\picplace{1cm}
\caption[7]{Distribution of normal (circles) and interacting 
(solid triangles) galaxies in the plane HI mass-to-blue luminosity
ratio $-$ scale height (a) and HI mass-to-blue luminosity ratio
$-$ $h/z_{0}$ (b).}
\end{figure}

 It is interesting to compare our observational results to the predictions
of N-body simulations. Quinn et al (1993) have addressed the specific 
problem of disk heating by small satellites, through stellar tree-code
calculations. They found that the disk heats vertically as well as radially;
however the radial spread is accompanied by a central mass concentration
and the inner disk scalelength $h$ decreases, while the central brightness
increases, and the scaleheight $z_0$ increases during an interaction.
  This corresponds quite well to our observations of a lower $h/z_0$ ratio
for interacting galaxies.
Quantitatively, they found in average a decrease of $h$ by
20\%, while $z_0$ increases by a factor 2. 
 Their simulations however over-estimate the disk heating, since their
dark matter halo is rigid, and cannot acquire energy and angular momentum,
and they ignore the gaseous component, that can dissipate away the energy
through radiation. They show that the internal heating of the primary
disk depends strongly on the mass concentration of the satellite: 
when the latter is denser than the primary, it is not disrupted until
the final merging, while when the satellite is more diffuse, it is
stripped all along the interaction, and the stripped particles can take
the heating away. In this case, about 95\% of the disk heating must occur
in the vertical direction, since the planar kinetic energy resides in 
the stripped stars of the satellite.  Indirect effects can then modify
somewhat this picture: if a strong spiral structure is triggered in the 
disk, this gravitational instability heats the disk, which spreads out
as the angular momentum is transferred outwards. The final result
depends however on the amount of gas present, and the Quinn et al (1993) 
simulations did not include it. 

Quinn et al (1993) predicted that the final scale height $z_0$ should
increase with radius, i.e. the thickened disk should flare. This is not
generally seen in our observations; only the double system VV 490 
reveals a strong flare (by 50\%); but the phenomenon in the simulations is
significant only after 10 kpc in radius, and our observations 
are not sensitive enough at large radii. 
Shaw \& Gilmore (1989, 1990) also found a constant scaleheight
for their sample of isolated undistrubed edge-on galaxies. This does
not seem to support the numerical predictions, since long after an 
interaction, the flaring is expected to subsist in the thick disk. 
Only in rare cases
the stellar disk reveals a flaring (e.g. NGC 3115, Cappacioli et al 1988).

 The simulations by Walker et al (1996) improved over the first work by
Quinn et al (1993) in treating the dark haloes consistently, and 
dealing with an order of magnitude more particles. However the results
are quite similar, they found a decrease of the scalelength of the inner disk
$h$ of about 15\%, while the envelope in the outer parts was enriched
through radial spreading of the primary disk. The thickness of the disk
increased by 60\% at the solar distance from the center. This figures
imply a disk heating only slightly inferior to what was found by Quinn 
et al (1993), and completely within the variations expected from
the nature of the satellite (dense or diffuse).  They also ignored
the gas component, which could bring more qualitative differences.
 Hernquist \& Mihos (1995) simulated minor mergers between gas-rich disks
and less massive dwarf galaxies. The large difference they found with
comparable simulations without gas, is the huge central mass concentration
driven by tidal torques due to tidally-induced bars on the gas. Half of
the gas mass is driven to a region less than 1 kpc across, and
triggers a starburst there. Consequent to the central mass concentration, 
the inner scalelength of the disk decreases, while the disk thickens. 
 Unfortunately, star-formation was not included in these simulations,
which might over-estimate the gas inflow. 

\section{Conclusions}
 We have studied a sample of 24 edge-on interacting galaxies and compared
them to other edge-on isolated galaxies, to investigate the effects
of tidal interaction on disk thickening. We find for most of the galaxies
a constant (within 20\%)
scaleheight with radius. Only one system (VV 490) reveals
a significant flaring. The average scaleheight $z_0$ is higher for
the interacting sample, and the scalelength is also smaller, so that 
the $h/z_0$ ratio is 1.5-2 times smaller than in isolated galaxies.
  This corresponds quite well to the predictions of N-body simulations
(Quinn et al 1993, Walker et al 1996): the gravity torques induced by
the tidal interaction produce a central mass concentration, while 
the outer disk spreads out radially, leading to a decrease of $h$.
 Most of the heating is expected to be vertical, since the planar
heating is taken away by the stripped stars either in the primary or
in the satellite. The quantitative agreement between observations
and simulations is rather good, given the large dispersion expected due to the
initial morphology of the interacting galaxies: a dense satellite 
will produce much more heating than a diffuse one, where stripped stars
take the orbital energy away; a mass-condensed primary will inhibit
tidally-induced spiral and bar perturbations, that are the source of heating
both radially and vertically. 

The fact that tidal interactions and minor mergers must have involved 
every galaxy in a Hubble time, and therefore also the presently 
isolated and undisturbed galaxies, tells us that the lower values
of $h/z_0$ observed for the interacting sample must be transient.  
Radial gas inflow induced by the interaction may have contributed to 
reform a thin young stellar disk, while the vertical thickening
have formed the thick disk components now observed in the Milky
Way and many nearby galaxies (e.g. Burstein 1979, Shaw \& Gilmore 1989). 
This process might be occuring all
along the interaction, so that the galaxy is never observed without
a thin disk. One cannot therefore date back the period of the
last interaction by the age of the thin disk, as has been proposed
by Toth \& Ostriker (1992) and Quinn et al (1993). The Milky Way,
experiencing now interactions with the Magellanic Clouds and a few
dwarf spheroidal companions, has still a substantial gaseous and 
stellar thin disk. More self-consistent simulations, including gas
and star-formation, must be performed to derive more significant 
predictions.

\acknowledgements{
VR acknowledges support from French Ministere de la Recherche
et de la Technologie during his stay in Paris.
This research has made use of the NASA/IPAC Extragalactic
Database (NED) and of the Lyon-Meudon Extragalactic Database
(LEDA) supplied by the LEDA team at the CRAL-Observatoire de
Lyon (France).}


\begin{thebibliography}{}
\bibitem{} Barnes J.E., 1988, ApJ 331, 699
\bibitem{} Barteldrees A., Dettmar R.-J., 1994, A\&AS 103, 475
\bibitem{} Bottema R., 1993, A\&A 275, 16
\bibitem{} Braine J., Combes F., 1993, A\&A 269, 7  
\bibitem{} Burstein D., 1979, ApJ 234, 829
\bibitem{} Capaccioli M., Vietri M., Held E.V., 1988, MNRAS 234, 335
\bibitem{} Chiappini C., Matteucci F., Gratton R., 1997, ApJ, in press
\bibitem{} Combes F., 1996, in "Barred galaxies and circumnuclear activity",
Nobel Symposium 98, ed. A. Sandqvist \& P.O. Lindblad, Lecture Notes in
Physics, Springer, vol. 474, 101
\bibitem{} de Grijs R., van der Kruit P.C., 1996, A\&AS 117, 19
\bibitem{} de Vaucouleurs G., de Vaucouleurs A., Corwin H.G. et al.,
1991, ``Third Reference Catalogue of Bright Galaxies'' (RC3),
Springer-Verlag
\bibitem{} Dickey J.M., Hanson M.M., Helou G., 1990, ApJ 352, 522
\bibitem{} Freeman K.C., 1970, ApJ 160, 811
\bibitem{} Frenk C.S., White S.D.M., Davis M., Efstathiou G., 1988, 
            ApJ 327, 507
\bibitem{} Gilmore G., Reid I.N., 1983, MNRAS 202, 1025
\bibitem{} Gunn J.E., 1987, in "Nearly Normal Galaxies", ed S.M. Faber,
           Springer, New York, p. 459
\bibitem{} Guthrie B.N.G., 1992, A\&AS 93, 255
\bibitem{} Hernquist L., Mihos J.C., 1995, ApJ 448, 41
\bibitem{} Huchtmeier W.K., Richter O.-G., 1989, A General
Catalog of HI Observations of Galaxies, Springer-Verlag, New York 
\bibitem{} Ibata R.A., Gilmore G., Irwin M.J., 1994, Nature 370, 194
\bibitem{} Karachentsev I.D., Karachentseva V.E., Parnovsky S.L., 1993,
Astron. Nachr. 314, 97
\bibitem{} Lin D.N.C., Pringle J.E., 1987, ApJ 320, L87
\bibitem{} Nieto J.-L., McClure R., Fletcher J.M. et al., 1990,
A\&A 235, L17
\bibitem{} Ostriker J.P., Tremaine S., 1975, ApJ 202, L113
\bibitem{} Pardi M.C., Ferrini F., Matteucci F., 1995, ApJ 444, 207
\bibitem{} Quinn P.J., Hernquist L., Fullagar D.P., 1993, ApJ 403, 74
\bibitem{} Reid N., Majewski S.R., 1993, ApJ 409, 635
\bibitem{} Reshetnikov V., Combes F., 1996, A\&AS 116, 417 (Paper I)
\bibitem{} Reshetnikov V.P., Hagen-Thorn V.A., Yakovleva V.A., 1993, A\&A
            278, 351
\bibitem{} Reshetnikov V.P., Hagen-Thorn V.A., Yakovleva V.A., 1996,
A\&A 314, 729
\bibitem{} Robin A.C., Haywood M., Creze M., Ojha D.K., Bienaym\'e O.,
1996, A\&A 305, 125
\bibitem{} Sackett P.D., 1997, ApJ, in press 
\bibitem{} Sandage A., Fouts G., 1987, AJ 92, 74
\bibitem{} Searle L., Zinn R., 1978, ApJ 225, 357
\bibitem{} Shaw M.A., Gilmore G., 1989, MNRAS 237, 903
\bibitem{} Shaw M.A., Gilmore G., 1990, MNRAS 242, 59
\bibitem{} Schweizer, F., 1990, in "Dynamics and Interactions of Galaxies",    
ed. R. Wielen, Springer p. 60
\bibitem{} Schweizer F., Seitzer P., 1988, Ap.J. 328, 88
\bibitem{} Schweizer F., Seitzer P., 1992, A.J. 104, 1039.
\bibitem{} Toomre A., Toomre J., 1972, Ap.J. 405, 142
\bibitem{} Toomre A., 1977, in "The evolution of galaxies and
stellar populations", ed . B. Tinsley \& R. Larson, New Haven, Yale 
University Press, p. 401
\bibitem{} Toth G., Ostriker J.P., 1992, ApJ 389, 5
\bibitem{} van der Kruit P.C., Searle L., 1981a, A\&A 95, 105
\bibitem{} van der Kruit P.C., Searle L., 1981b, A\&A 95, 116
\bibitem{} van der Kruit P.C., Searle L., 1982a, A\&A 110, 61
\bibitem{} van der Kruit P.C., Searle L., 1982b, A\&A 110, 79
\bibitem{} van der Kruit P.C., 1988, A\&A 192, 117
\bibitem{} Walker I.R., Mihos J.C., Hernquist L., 1996, ApJ 460, 121

\end{thebibliography}
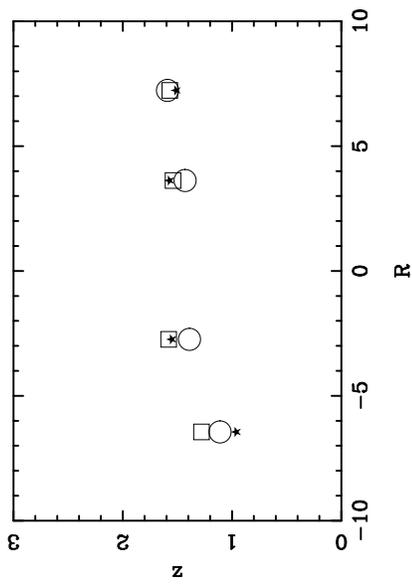
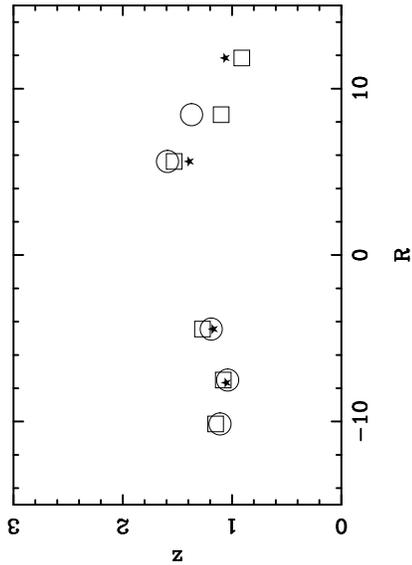
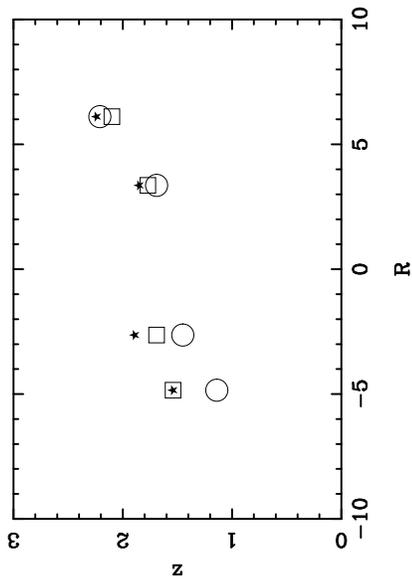
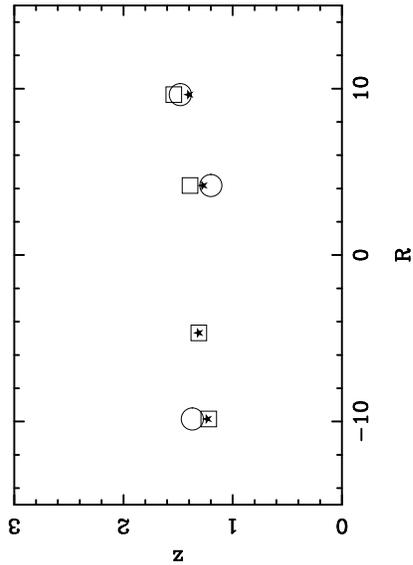
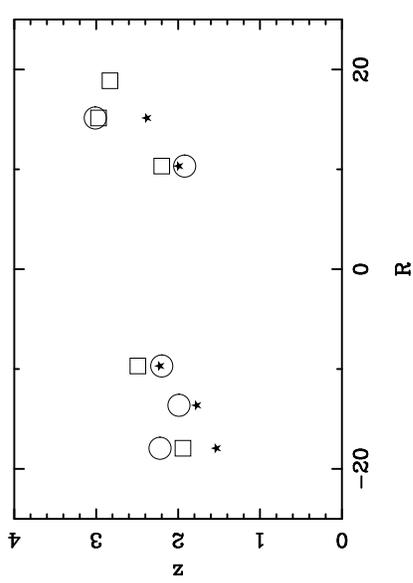
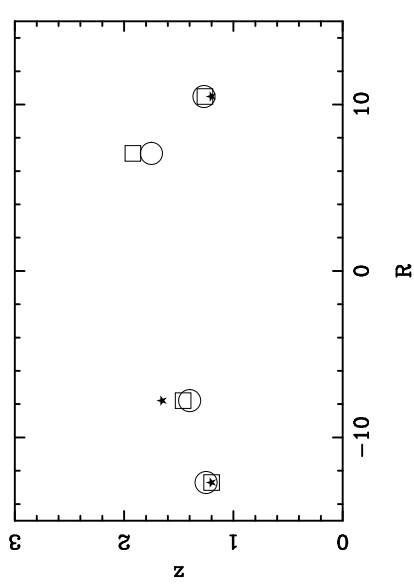
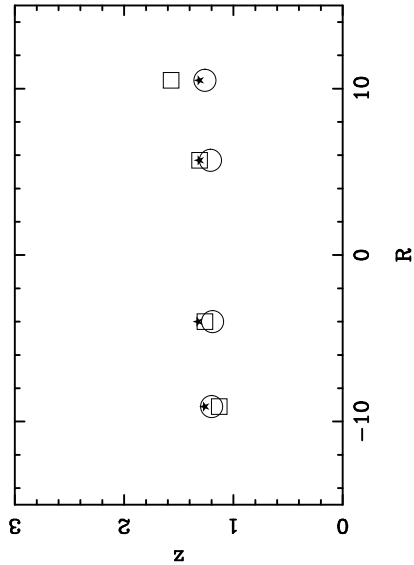
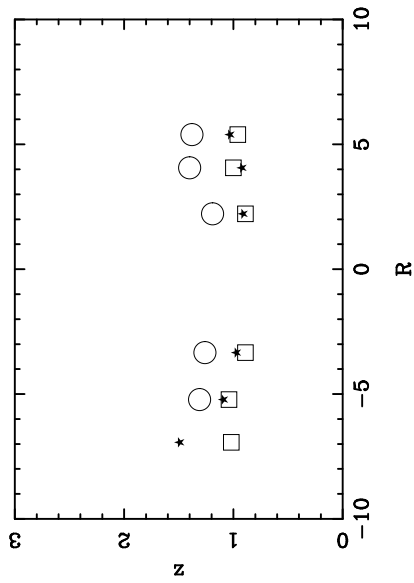
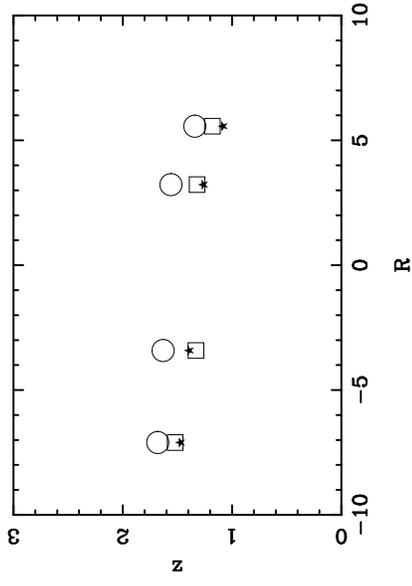
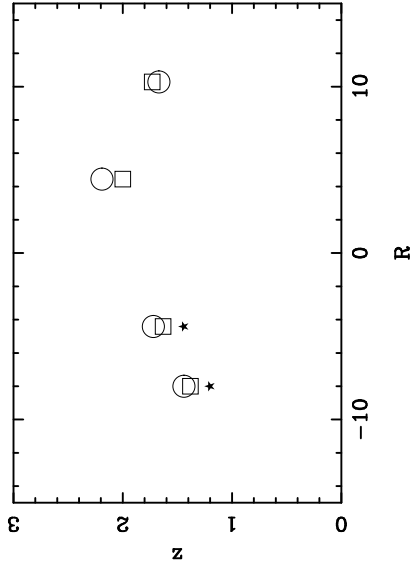
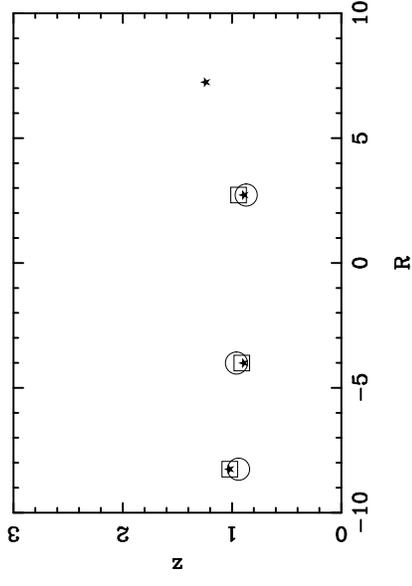
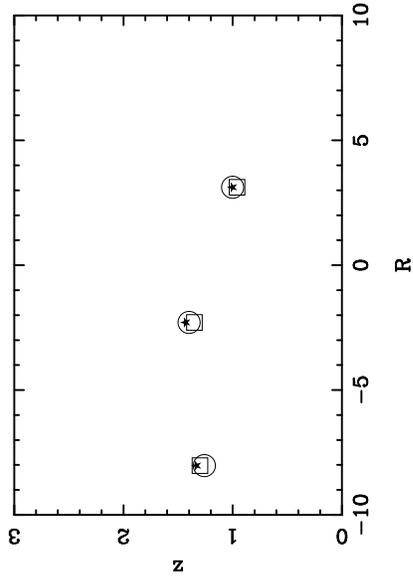
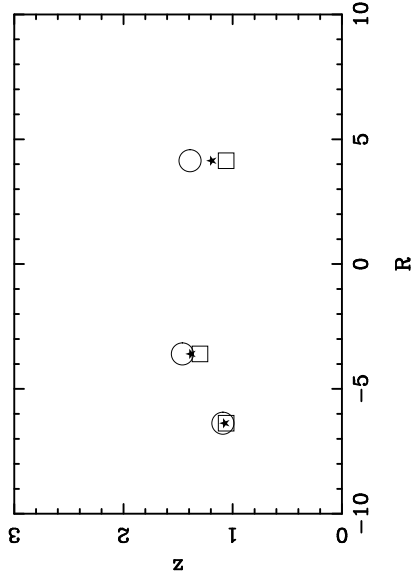
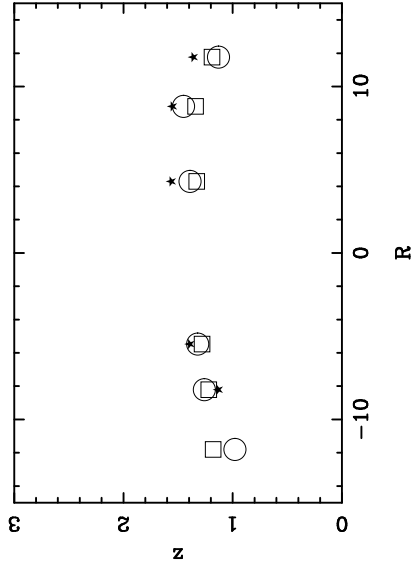
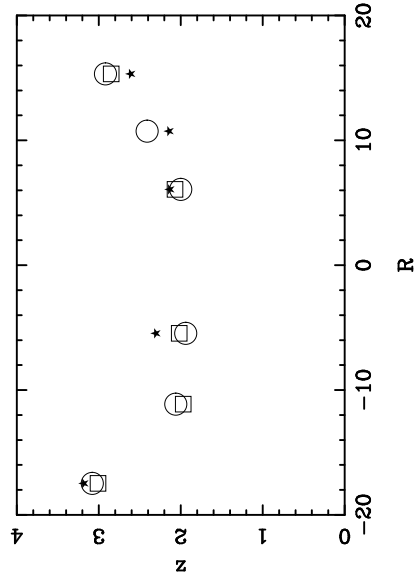
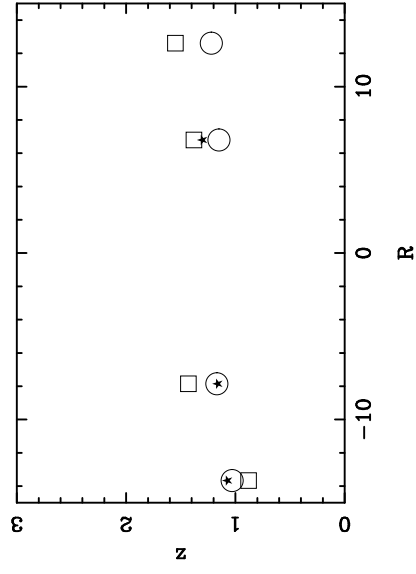
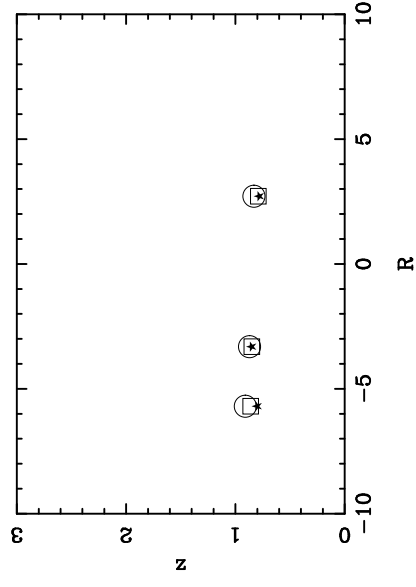
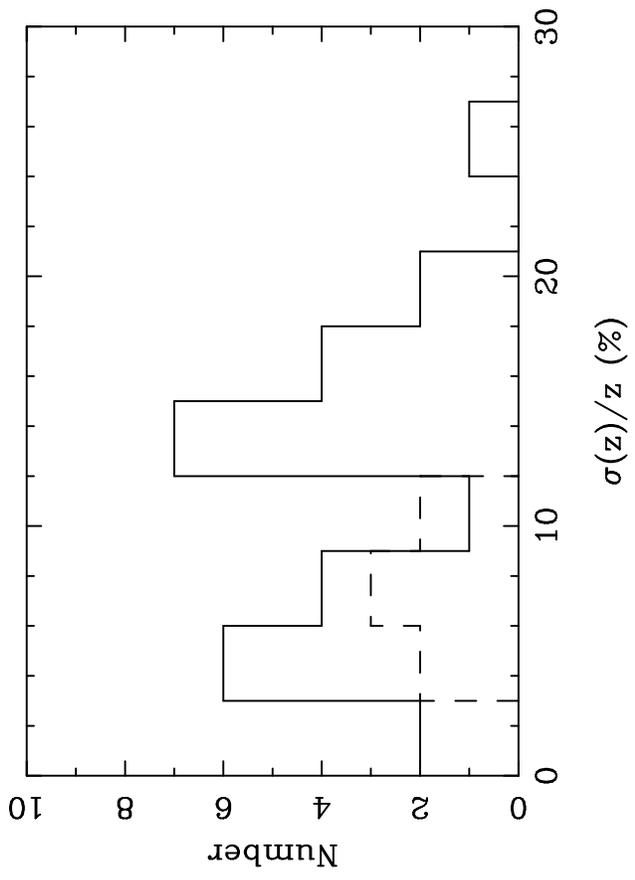
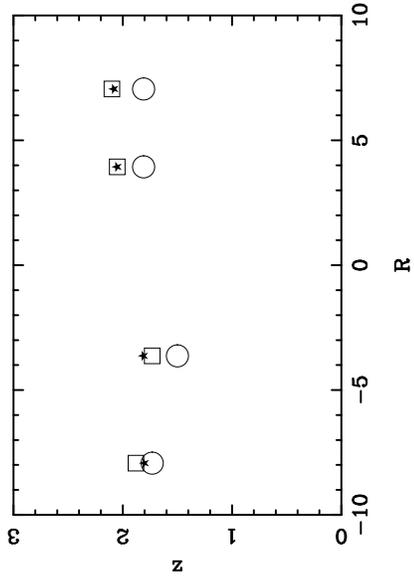
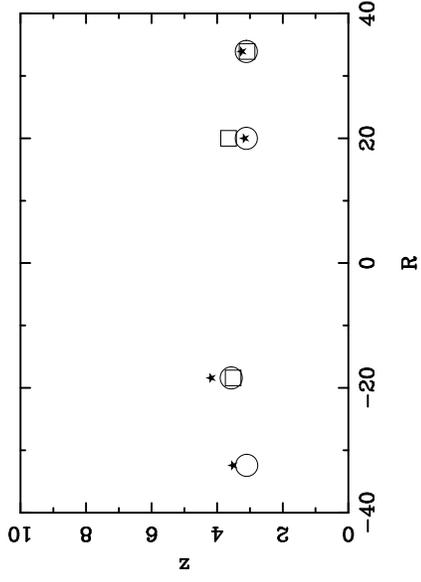
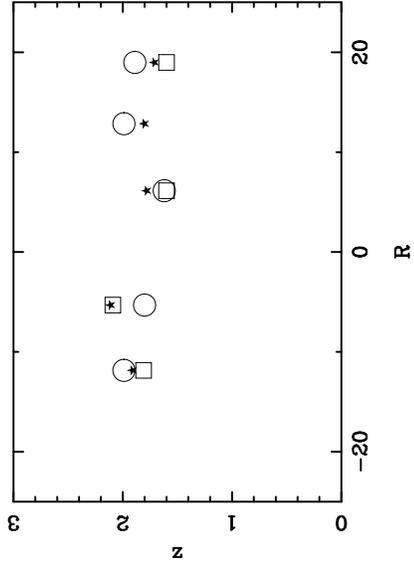
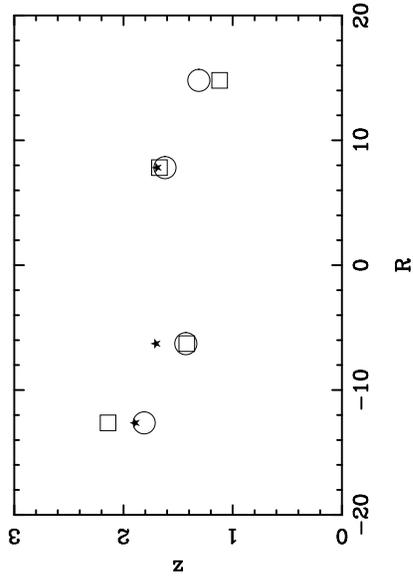
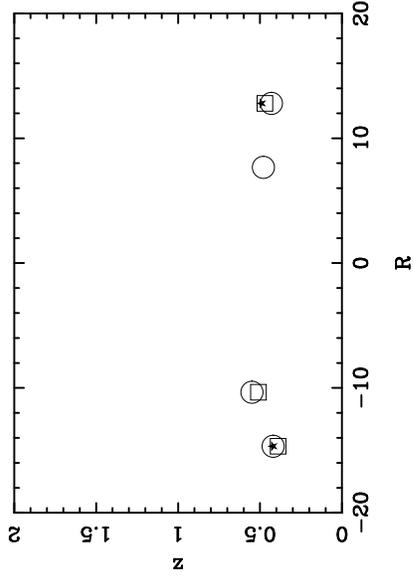
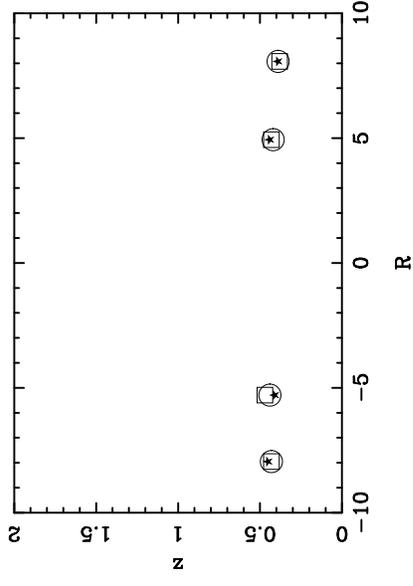
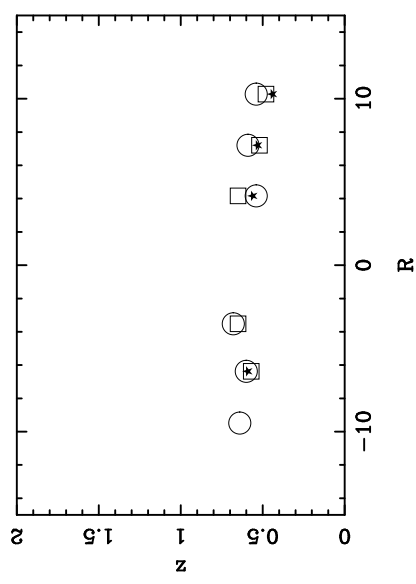
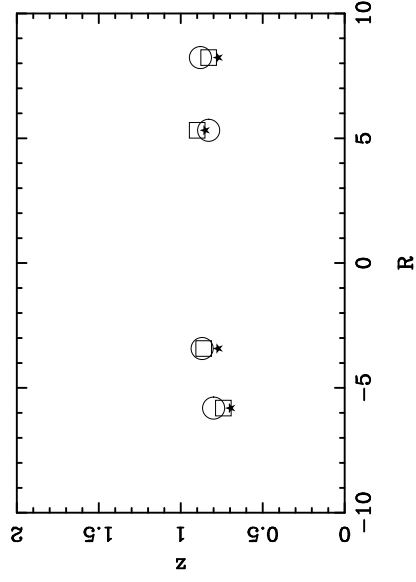
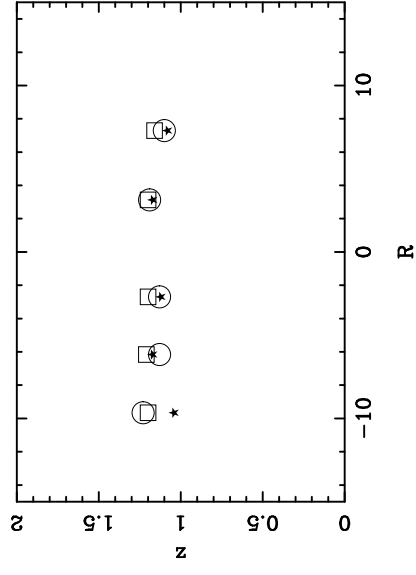
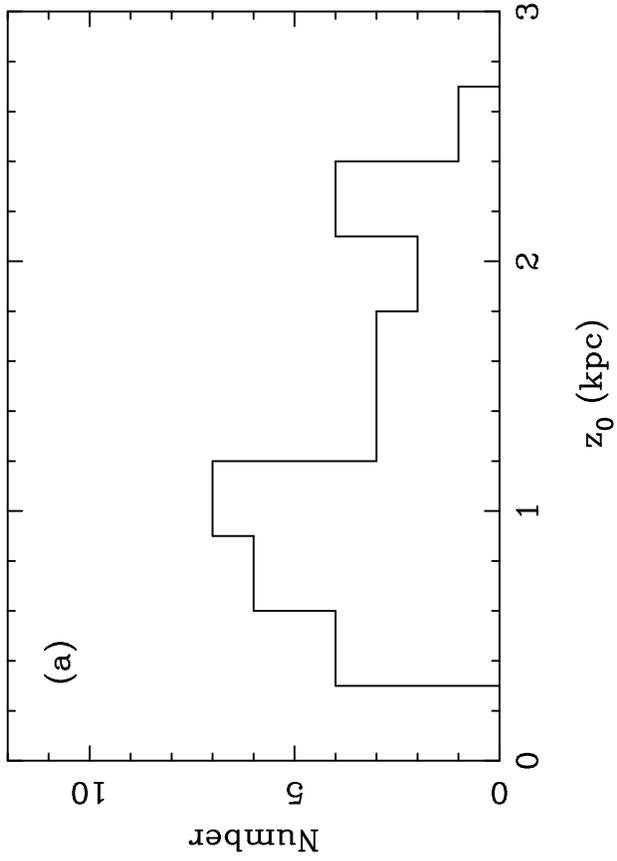
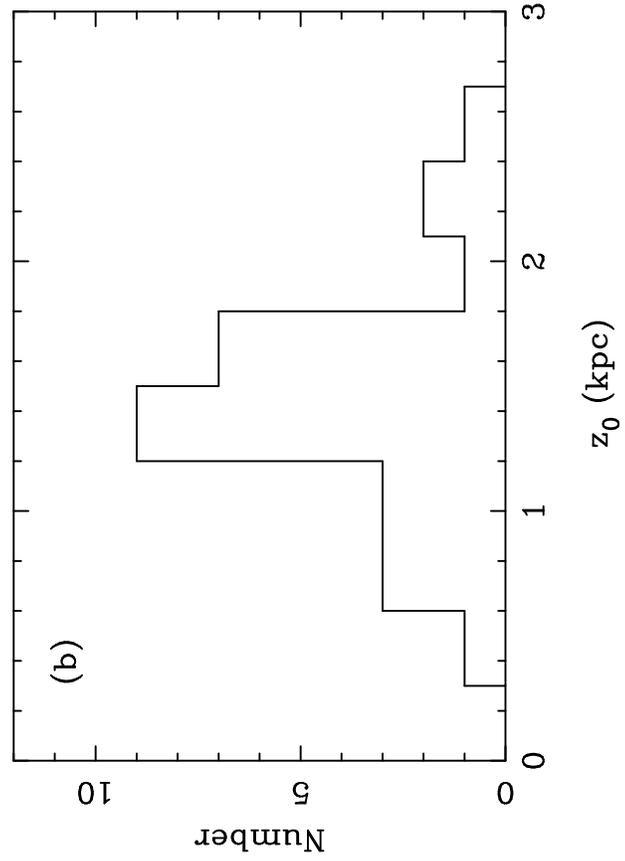
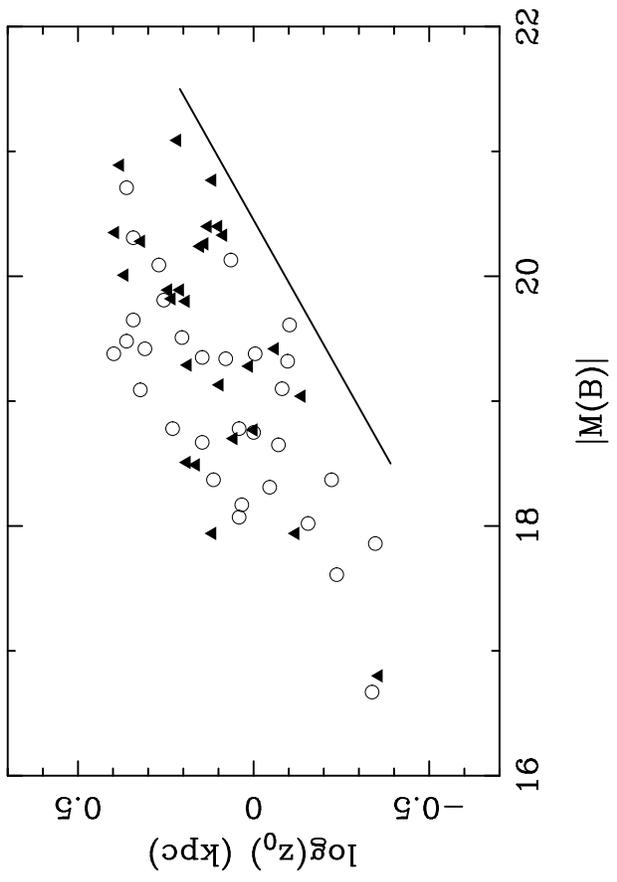
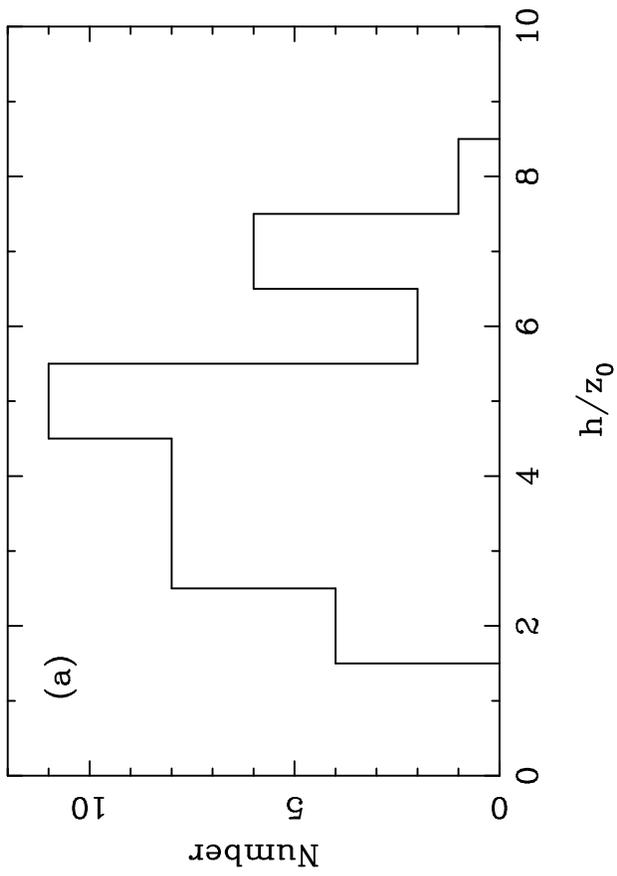
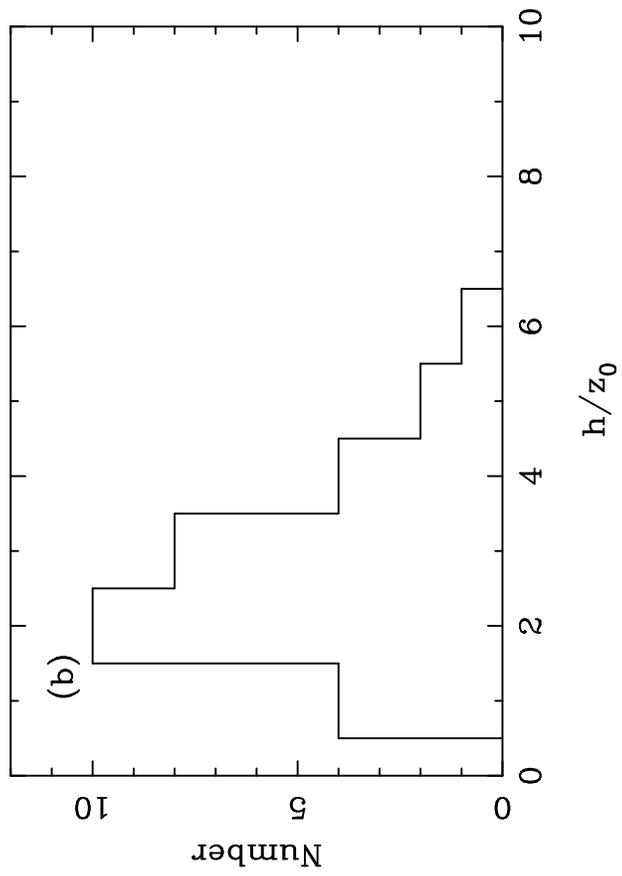
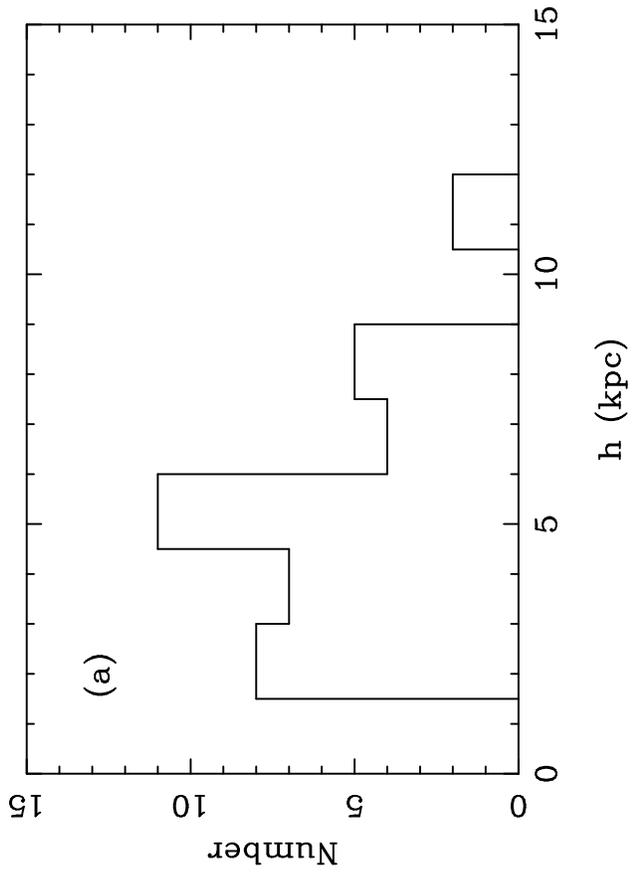
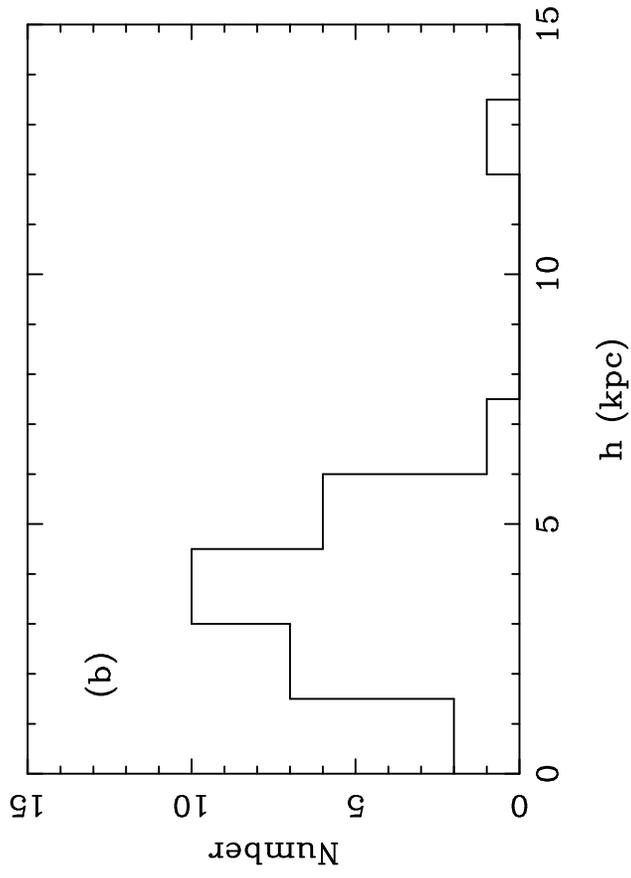
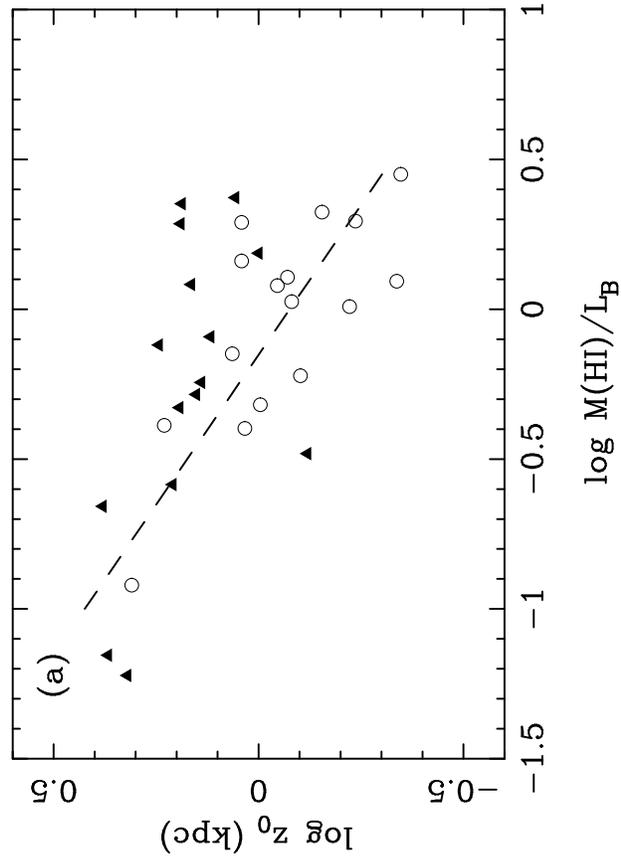
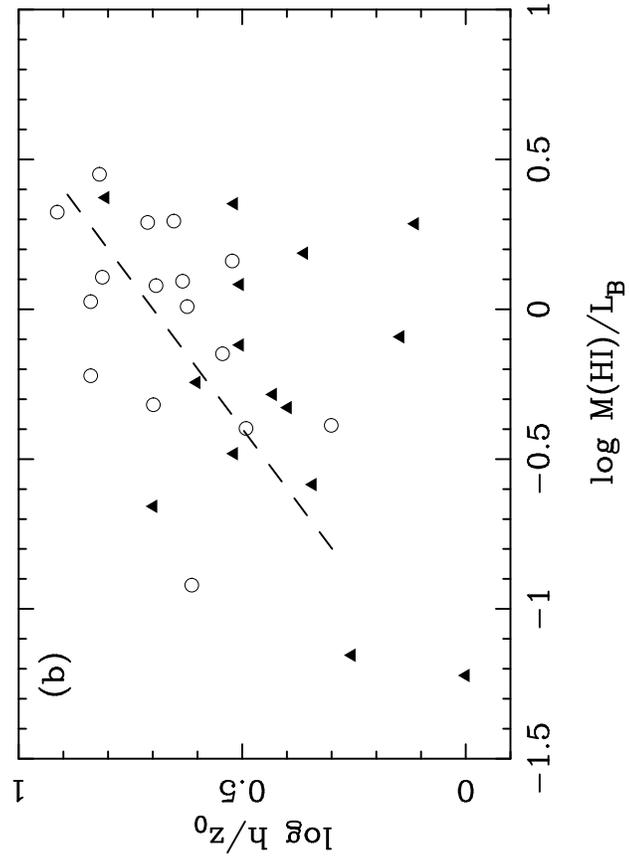
\end{document}